\documentstyle[12pt]{article}
\begin{document}
\newcommand{\be}{\begin{eqnarray}}
\newcommand{\ee}{\end{eqnarray}}
\title{Gluon evolution at low $x$ and the longitudinal structure function}
\author{Ranjita Deka$^1$,D.K.Choudhury$^2$\\
$^1$Department of Physics,Pragjyotish 
College,Guwahati-781009,India.\\
$^2$ Department of Physics, Gauhati University, Guwahati- 781014,India. }
\maketitle

\abstract{}
 We obtain an approximate analytical form of the gluon
distribution using the GLAP equation with a factorization ansatz,and test
its validity by comparing it with that of Gluck,Reya and Vogt at low $x$
regime.  We also present calculations of the longitudinal structure
functions. 

\section{Introduction}
    In deep inelastic scattering we can directly study the structure of the proton,
particularly the parton distributions \cite{reya,ap,robs}. The
perturbative QCD gives the $Q^2$ evolution and asymptotic limits of the
structure function.More recently, the study of structure functions at low
$x$ \cite{zeus1,akm,kms,vio,mrs,grv,ball1,ball2} has become topical in
view of the
 high energy collider like HERA  \cite{zeus2,h1}
where previously unexplored small $x$ regime is being reached.In the small $x$
 regime 
gluons are expected to be directly measurable. This expectation has led to
several 
approximate phenomenological schemes  \cite{coop,cdl,prytz1,prytz2,lanc}.
Specifically
 measurement 
of longitudinal structure function $F_L$ has long been advocated
\cite{coop,cdl,orr,neerven,meng,blum}
as a direct probe of the gluon density at small $x$.There is
even helpful
suggestion  \cite{coop,cdl,meng} that precise measurement of 
 $F_L$ should indeed be possible at HERA regime $x \leq 10^{-2}$ and
$ Q^2 \sim 10 \sim 100 GeV^2$.

     The present paper aims at obtaining an approximate analytical form of gluon
distribution using GLAP equations \cite{glap,abbot,grib}.An additive
assumption on the way 
to this 
is the factorization of the $x$ and $t ( t = ln Q^2/\Lambda^2)$ dependence of the
gluon
density. We test the validity of the assumption by comparing with the leading gluon 
density of Gluck, Reya and Vogt (LO-GRV) \cite{grv2}. 
We also use our results to calculate $F_L(x,Q^2)$ using its relation with gluons
 \cite{coop,cdl} within the range of validity of our assumption and compare with those
of collinear  \cite{neerven} and $ K_T $factorisation  \cite{blum} approaches.


\section{$x$  and $t$ evolution of the gluons}.

 We start our derivation,taking only the leading term of the gluonic kernel of the 
 GLAP equations  \cite{glap,abbot,grib}

\be                           
\frac{\partial G(x,t)}{\partial t}=\frac{3\alpha_s(t)}{\pi}\left[\left\{\left(\frac{11}
{12}-\frac{N_f}{18}\right)+ln(1-x)\right\}G(x,t) \right.
\nonumber\\
\left .+\int_x^1 dx\left(\frac{zG(x/z,t)}{1-z}
-\frac{G(x,t)}{1-z}+\left(z(1-z)+\frac{(1-z)}{z}\right) G(x/z,t)\right)\right]
\ee

where $ G(x,t) = xg(x,t), \alpha _s(Q^2)=\frac{12\pi}{33-2N_f} log(Q^2/\Lambda^2)$
and $N_f=$no. of flavours. 
Here we have neglected the contribution of the singlet structure function as it is 
expected to be small in the low $x$ regime.In order to facilitate our analytical 
solution, let us assume that the $x$ and $t$ dependence of the structure function
are 
factorizable  \cite{pri,dkc}

\be
G(x,t)=g(x)h(t)
\ee

with the condition

\be
g(x)=G(x,t_o)
\ee

so that

\be
g(x)\frac{\partial h(t)}{\partial t}=\frac{3\alpha_s(t)}{\pi}\left[\left\{\left(\frac{11}
{12}-\frac{N_f}{18}\right)+ln(1-x)\right\} g(x)h(t)\right .
\nonumber\\
\left . 
+\int_x^1 dz\left(
                  \left\{\frac{zg(x/z)-g(x)}{1-z}\right\} h(t)
             \right)
+\left(z(1-z) + \frac{1-z}{z}\right)g\left( \frac{x} {z}\right) h(t)
\right]
\ee

Dividing by $g(x)$ throughout we have

\be
\frac{\partial h(t)}{\partial t}=\frac{3\alpha_s(t)
h(t)}{\pi}\left[\left\{\left(\frac{11}{12}-
\frac{N_f}{18}\right)+ln(1-x)\right\} 
\right. 
\nonumber\\ 
\left.
+\int_x^1 dz\left\{\frac{zg(x/z)-g(x)}{(1-z) g(x)}
+\left(z(1-z)+\frac{(1-z)}{z}\right)\frac{g(x/z)}{g(x)}\right\}\right]
\ee

or

\be
\frac{\partial h(t)}{h(t)}=\frac{3\alpha_s(t)\partial t}{\pi}
\left[\left\{\left(\frac{11}{12}-\frac{N_f}{18}\right)+ln(1-x)\right\}+I_g(x)\right]
\ee

where

\be
I_g(x)=\int_x^1
dz\left\{\frac{zg(x/z)-g(x)}{(1-z)g(x)}+\left(z(1-z)+\frac{1-z}{z}\right)
\frac{g(x/z)}{g(x)}\right\}.
\ee

using eq.(2) and solving eq.(6) we find

\be
\ln h(t)=ln
t^{\left[\frac{36}{25}\left\{\left(\frac{11}{12}-\frac{N_f}{18}\right)+ln(1-x)+
I_g(x)\right\}\right]}
\ee

or
\be
h(t)=t^{\left[\frac{36}{25}\left\{\left(\frac{11}{12}-\frac{N_f}{18}\right)+\ ln(1-x)+
I_g(x)\right\}\right]}.
\ee

Therefore,

\be
G(x,t)=G(x,t_o)\left(\frac{t}{t_o}\right)^{\left[{36\over25}\left\{\left(\frac{11}{12}-
\frac{N_f}{18} \right)+ \ln(1-x)+I_g(x)\right\}\right]}
\ee

where

\be
I_g(x)=\int_x^1 dz\left[\frac{zG(x/z,t_0)-G(x,t_o)}{(1-z)G(x,t_o)}
+\left(z\left(1-z\right)+\frac{1-z}{z}\right)\frac{G(x/z,t_o)}{G(x,t_o)}\right].
\ee

Knowing the input parametrisation of the gluons and evaluating $I_g (x)$
numerically,we can
find the gluon density for various $x$ and $t$ using eq.(10) .We note that in the
limit 
$x \rightarrow 0 $,eq.(10)  has the universal limiting behaviour

\be
G(x,t) = G(x,t_o)\left(\frac{t}{t_o}\right)^{ \frac{36}{25}\ ln(1/x)}
\ee

to be compared with the standard double leading logarithmic expectations 
 \cite{rujula,mrs2}

\be
G(x,t) \sim  \exp\left[ \ ln\frac{1}{x} \ ln t\right]^{ \frac{1}{2}}
\ee

which is not factorizable in $ x $ and $ t $ ,while log G(x,t) is factorizable.

\section{The longitudinal structure function $ F_L(x,Q^2) $ }

Measurement of $F_L(x,Q^2)$ at low $x$ have been used to extract the gluon density
 \cite{coop,cdl}

\be
xG(x,Q^2)=\frac{3}{5}5.8\left[\frac{3\pi}{4\alpha_s}F_L(0.417x,Q^2)
- \frac{1}{1.97}F_2(0.75x,Q^2)\right]
\ee

for four active flavours. At low values of $x$, the gluon contribution
dominates and to a fair approximation,

\be
F_L(ax,Q^2) \cong   \frac{2\alpha_s}{3\pi}\frac{1}{1.74} xG(x,Q^2)
\ee

Here $\alpha_s$ is the QCD coupling strength and $a$ is a parameter whose 
value is $0.417$ for $F_L$  \cite{cdl}.
Using eq.(10) in eq.(15) we can thus obtain the longitudinal structure
function.
The behaviour of $F_L$ is known in $O(\alpha_s^2)$  \cite{neerven}
in collinear approach and was also studied in $O(\alpha_s)$ within $K_T$ 
factorization scheme in \cite{blum} .In our analysis, we compare our
prediction for $F_L$ 
with those of \cite{neerven} and \cite{blum} and study their
differences.

\section{Results and conclusions}

 The factorization assumption eq.(2) is in general not valid in theoretical
 framework 
describing the scaling violation in QCD i.e. in LO Altarelli-Parisi equations.
Even in DLA only $\log G(x,t)$ is factorizable in $x$ and $t$. We have
 therefore
 attempted to see how the predictions with this assumption  compare with those
 of
 gluon distribution which does not have such an assumption, like
 LO-GRV  \cite{grv2}.
       This
 will enable us to find the kinematical region of its approximate validity.

 In Fig. 1(a-l) we show the prediction of eq.(10)(curve marked 1) with
factorization
 ansatz eq.(2) and compare with LO-GRV \cite{grv2}(curve marked 2) for
representative $Q^2$
 values $4.5$, $6$, $8.5$, $10$, $20$, $40$, $80$, $100$, $160$, $1600$, $10^4$,
 and $10^5$ GeV$^2$ and $ 10^{-4} < x < 10^{-1}$ starting with the
evolution at $Q_o^2 =
 4$ GeV$^2$. These figures show the following feature for smaller $x$ range 
$(x < 10^{-2}):$
 at fixed  $x$ ,the difference between the two  increases as $Q^2$ is increased.As an
 illustration, at $x \sim 10^{-2}$ the difference increases     
from  $\sim 0.1\% $ to $20\%$ as $Q^2$ increases from $4.5$ to $160$ GeV$^2$. 
For each $Q^2$,
there is a cross-over point for both the curves,where both the predictions are
 numerically equal.The cross-over point shifts to lower $x$ as $Q^2$  increases .
Approximately,such cross-over occurs between $10^{-2} < x <10^{-1}$ for 
$Q^2 \sim 4.5 - 160$
 GeV$^2$ and between $10^{-3} < x < 10^{-2} $ for $Q^2 
\sim 160 - 10^5 $ GeV$^2$. We can therefore find the limited range of $x$ and $Q^2$ 
where our
 approximate expression for gluon density differs from LO-GRV by not more than $20\%$
 as shown in Fig.$2$.
 
 In Fig.$3$ we compare our result for $F_L$ with those obtained with collinear
\cite{neerven} and $K_T$ factorization approach \cite{blum},at $Q^2
= 20 $ GeV$^2$. Our result
is found to be higher than those of \cite{neerven} and \cite{blum}  . As
an illustration at $x \sim 10^{-2}$
 our result differs from \cite{neerven} by $33\%$,$ 67\%, $ and
$66\%$, 
corresponding 
to full $F_L(O(\alpha_s^2))$, $O(\alpha_s)$ and $O(\alpha_s^2)$ respectively. On
 the other hand it differs by $98\%$ with \cite{blum} .The difference
increases as $x$ decreases.
However , as the cross-over of the gluon distribution eq.(10)  with LO-GRV
occurs in
 the range $ x \sim 10^{-1} - 10^{-2} $ for $Q^2 \sim 20 $ GeV$^2$,the 
prediction may not be reliable
 for $ x < 10^{-2} $. It however calls for  quantitative study of 
$O(\alpha_s^2)$ and quark contributions within the present approach.
 
 To conclude we have shown that for a limited range of $x$ and $Q^2$,the gluon
density eq.(10) with factorization is numerically equivalent to LO-GRV.

 We have then predicted the longitudinal structure function $F_L$ within that
range and compared with those obtained in other approaches \cite{neerven,blum}.
Our result is 
found to be higher than those of \cite{neerven,blum}.It will be interesting to see
how our
prediction for $F_L$ compares with the results of forthcoming experiments at HERA.

\it Acknowledgements.

Both the authors would like to thank Dr.A.Saikia for useful discussions on gluons.
One of the authors (R.D)is  grateful to A.M.Cooper-Sarkar for giving suggestions
on longitudinal structure functions.

\end{document}